\documentclass[runningheads]{llncs}

\usepackage{url}
\usepackage{amssymb}
\usepackage{amsmath}
\usepackage{float}
\usepackage{graphicx}
\usepackage{epsfig}
\usepackage{caption}
\usepackage{subcaption}
\usepackage{subfig}

\urldef{\mailsa}\path|{ateniese, hitaj, mancini, verde, villani}@di.uniroma1.it, |

\begin{document}
\title{No Place to Hide that Bytes won't Reveal:\\ Sniffing Location-Based Encrypted Traffic to Track a User's Position}
\titlerunning{Sniffing LBS Encrypted Traffic to Track a User's Position}
\author{Giuseppe Ateniese\and Briland Hitaj\and Luigi V. Mancini\and Nino V. Verde\and Antonio Villani}
\authorrunning{G. Ateniese\and B. Hitaj\and L.V. Mancini\and N.V. Verde\and A.Villani}
\institute{Dipartimento di Informatica, Universit\`a di Roma ``La Sapienza'',\\Via Salaria 113, 00198 Rome, Italy\\ \mailsa\\ \url{http://www.di.uniroma1.it/}}

\maketitle

\begin{abstract}
News reports of the last few years indicated that several intelligence agencies are able to monitor large networks or entire portions of the Internet backbone. Such a powerful adversary has only recently been considered by the academic literature. \\
In this paper, we propose a new adversary model for Location Based Services (LBSs). The model takes into account an unauthorized third party, different from the LBS provider itself, that wants to infer the location and monitor the movements of a LBS user. We show that such an adversary can extrapolate the position of a target user by just analyzing the size and the timing of the encrypted traffic exchanged between that user and the LBS provider. We performed a thorough analysis of a widely deployed location based app that comes pre-installed with many Android devices: GoogleNow. The results are encouraging and highlight the importance of devising more effective countermeasures against powerful adversaries to preserve the privacy of LBS users. 
\keywords{location-based services, network traffic analysis, GoogleNow, privacy, mobile devices}
\end{abstract}

\section{Introduction}
Modern surveillance systems that track the movements of cellphone users are more sophisticated than ever. Intelligence agencies can easily locate the cell tower used by a target and find his location. According to a Washington Post article~\cite{WashingtonPost:Systems}, it is possible to exploits security vulnerabilities in the network used by mobile carriers around the world to provide services to their traveling customers. This network is called the Signaling System 7 (SS7) and once access to it is obtained, it is possible to track the location of anyone in the world and learn whether a person is walking down a specific street, driving, or taking a flight. When the approximate location of a target is known, {\em stingrays}~\cite{ArsTechnica:MachinesStealPhoneData} (or {\em fake} towers) can be used to redirect calls, monitor Internet traffic, steal phone's data, and even install malware. \\
\indent
These attacks, however,  are active and (partially) intrusive. In particular, they are not completely stealthy and leave traces. Indeed, queries to the SS7 network can be logged and phones could be configured by experts to detect stingrays (e.g., IMSI-Catcher Detector on Android phones). In addition, these attacks can only be carried out in cooperation with vendors, mobile carriers, or ISPs.  Often the target is in a foreign country and special permissions or agreements must be in place to be able to track his movements. \\
\indent
In this paper we show that it is possible to locate the position of a cellphone by simply monitoring the traffic of certain phone applications that provide location-based services (LBS). Clearly this is simple if the traffic is in the clear, but our main contribution is to show that it is possible to track users even when the traffic is properly encrypted. We believe our method will have significant implications in the way location-based services are provided. LBSs are often accessed through apps that will be referred to as \emph{Location Based Apps} or LBAs. LBAs are used to find friends and restaurants nearby, to locate points of interest, to check public transport timetables and even to search for deals or special offers. Several physical retailers (e.g., Best Buy, Kohl’s), also deploy location-based promotions to push notifications while the consumer is in or near the store. TripAdvisor, Booking.com and weather forecasting applications are other examples of LBAs. \\
\indent
It is difficult to protect the privacy of users while at the same time provide useful LBSs. It is possible to obfuscate the exact position of a user but these obfuscation techniques are rarely adopted by vendors (location data is too valuable to them). Moreover, customers appreciate services or information they receive and do not seem concerned about sharing their location data with LBS providers. \\
\indent
The contribution of our paper is to show that any third party can infer a user's position by just analyzing the encrypted traffic from that user to the LBS provider. This can be performed in a non-intrusive way, without leaving any traces. For instance, an intelligence organization could monitor routers belonging to some  Autonomous System (AS) traversed by LBS's packets and this would be enough to infer a target's position in a foreign country without involving that country's ISPs, mobile carriers, or any other local entities. Encryption or NAT'd addresses do not help much in this scenario. Indeed, we leverage results of previous works on analysis on encrypted traffic which already highlighted the possibility of identifying apps installed on a device \cite{Conti:2015:CYH:2699026.2699119,Stober:2013:YSY:2462096.2462099}, or the presence of a specific user within a network \cite{NinoVerdeNATleftBehind}. \\
\indent
Another important point to consider in this context is that current LBAs have started to adopt push technology solutions to send ``the right information at the right time''~\cite{GoogleNow:Website}. This is the maxim of the GoogleNow app which comes preinstalled on most Android devices and provides several services that are tied to the user's position. Several LBAs that come preinstalled on a phone do not even ask permission to use location data. The adoption of push technology implies that the user is continuously tracked by the LBS provider, even when the app runs in the background~\cite{CMU:Raffalink}.

\paragraph{Contributions.}
In this work we put forward a new privacy problem related to LBSs. We introduce a new adversary model for LBSs and propose a technique that unauthorized third parties may use to infer the position of a target user. Furthermore, we analyze one of the most popular LBAs, GoogleNow, and we show that the analysis of its encrypted network traffic reveals the position of a user with high accuracy. 
This research is inevitably controversial. The method we developed could be used to undetectably monitor movements of users and abuse their privacy rights. However, it should be considered as a warning to the research community to spur more research in the area and come up with effective countermeasures.

\paragraph{Organization.}
The rest of this paper is organized as follows. In Sect.~\ref{sec:AdversaryModel}, we introduce the adversary model. In Sect.~\ref{sec:dataanalysis}, we detail the different phases of the attack. In particular, Sect.~\ref{subsec:datacollection} explains the method used by the adversary to collect the relevant data from the LBS provider. Then, Sect.~\ref{subsec:statistical} details the data analysis approach that can be used to infer user locations. Sect.~\ref{sec:exp-and-res} and \ref{sec:areasize} respectively report on the results achieved when analyzing GoogleNow, and on the strategy to select the points that should be monitored by the attacker. In Sect.~\ref{sec:related}, we review previous work. Finally, in Sect.~\ref{sec:conclusions} we draw the conclusions and discuss some possible future works.

\section{The Adversary Model}
\label{sec:AdversaryModel}
We assume the existence of an adversary $\mathcal{A}$ that can sniff the network traffic of a mobile device. The adversary does not need to intercept the entire network traffic but just the packets that are exchanged between the LBA and the LBS provider. The adversary may do so by compromising one of the network devices of any AS that routes the information between the mobile device and the LBS. We assume that the adversary does not want to be detected, and therefore he does not compromise the mobile device nor change the content of network packets. 
It is well-known that the NSA can identify users around the world of specific services (such as TOR) by detecting  packet ``fingerprints'' and monitoring large portions of the Internet. NSA accomplishes this by collaborating mainly with US telecoms firms under various programs~\cite{NSA:TorAttack}. \\
\indent
We assume that the adversary is able to identify and isolate the network traffic of the user he is interested in, and, among those packets, he is able to identify and isolate packets that are generated by the LBA. The adversary is able to determine where discrete communications begin and end (such as the download of updated information from the LBS). This is possible, for example, by observing typical communication patterns of the LBA. Note that if the network traffic is not encrypted, then our adversary may trivially inspect the packet content and determine the location of the mobile device. Therefore, we assume that the network traffic exchanged between the mobile device and the LBS is encrypted via SSL/TLS. Furthermore, we assume that the LBS provider does not use any mechanisms to protect the privacy of its users, such as k-anonymity cloaking, etc. This assumption is based on the fact that current LBS providers do not implement these mechanisms. \\
\indent
To launch the traffic analysis attack that we consider in this paper, the adversary must build a knowledge base that summarizes the network traffic exchanged between the LBA and the LBS when the mobile device is located in certain locations of interest. We assume that the adversary can collect this data by using bogus accounts and virtual mobile devices.  

\subsection{Example Scenario}
\label{par:ex-scenario}

\begin{figure}
\centering
 \includegraphics[width=0.8\linewidth]{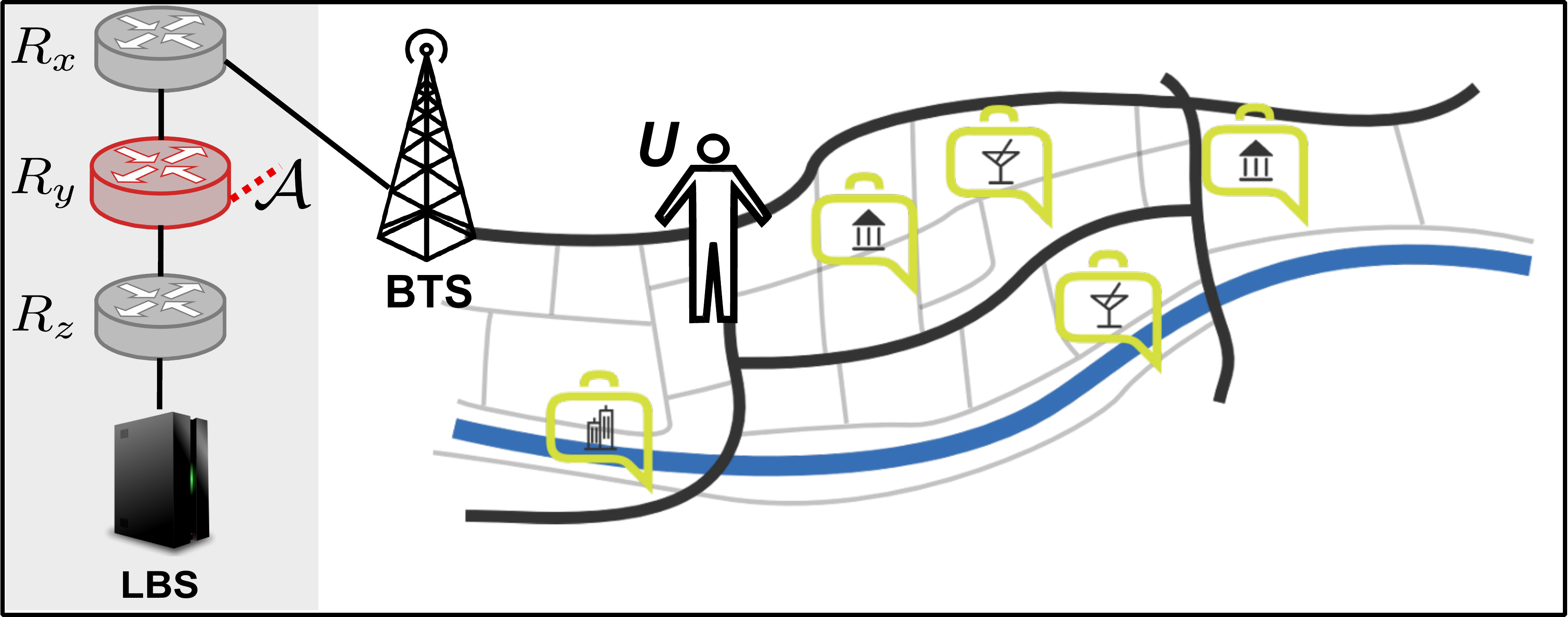}
 \caption{Attack scenario}
 \label{fig:scenario}
\end{figure}

Figure~\ref{fig:scenario} represents a possible attack scenario. The target $U$ is a user with a mobile device; the adversary is a malicious entity $\mathcal{A}$ that is after the geographical location of $U$.
In this scenario, $\mathcal{A}$ is not allowed to collude neither with the LBS nor with the ISP (the owner of the Base Transceiver Station - BTS) otherwise locating the user would be trivial (but intrusive) or even impossible if the LBS or the ISP refuse to provide private information. 
The LBS provider will keep sending new and relevant information to the device as the user moves around. This can happen without the user's intervention via push communication. Several LBAs, more prominently GoogleNow, work this way.
In this paper we analyze under which conditions collecting the encrypted traffic exchanged between the LBS provider and the device is enough to determine the exact user's geographical position. Figure~\ref{fig:scenario} depicts possible position in the network where $\mathcal{A}$ can sit to intercept the network traffic. This is represented by the router $R_y$ highlighted in red. However, it is worth mentioning that $\mathcal{A}$ can be potentially in any router laying in the path from the user's device toward the LBS. Furthermore, those routers may belong to different ASs. Thus, the adversary may infer the user position from ASs different from the one of the user. He can even perform the attack from foreign countries without involving mobile carriers, or any other local entity.

\section{The Attack}
\label{sec:dataanalysis}
In this section, we detail the approach used by the adversary to infer the actual position of a target user. The entire approach can be logically divided into two steps: the \emph{data collection} phase, and the \emph{candidate locations selection}. The aim of the \emph{data collection} phase is to collect enough information from the LBS provider to learn how different locations can be distinguished from each other. 
During this phase, the adversary builds up its knowledge base that will later be used to infer the most probable locations the target can be found in. For the sake of simplicity, we will assume that the LBS sends the same information to all users in the same location. Taking into account that the LBS may send personalized information to its users is left as future work.

\subsection{Data Collection}
\label{subsec:datacollection}
Suppose that the adversary is interested in localizing users in a given area. First, the adversary logically divides the entire area in $n$ subareas, and arbitrarily chooses a point in each subarea as a representative for that location. The size of the subareas is chosen according to the desired accuracy and the granularity of the information provided by the LBS. Hence, the adversary comes with a set of point locations $\mathcal{L}=\{l_1,l_2,\dots l_n\}$. \\
Then, the adversary collects data from the LBS about all the locations in $\mathcal{L}$. This can be accomplished by using the same LBA of regular users and by spoofing the GPS coordinates pretending to be in each location $l_i$ for $1\leq i\leq n$. The adversary periodically performs the  procedure above to learn the traffic pattern of the LBS over time (e.g., data sent by the LBS may change according to daily or weekly trends). \\
\indent
The network traffic that the adversary collects is used to build its knowledge base. 
The following steps are performed during this phase: 
\begin{description}
 \item[Prefiltering:] The network traffic is analyzed with a network protocol analyzer, and only the packets directed towards, or coming from, the network of the LBS are preserved.
 \item [Knowledge Base Record Composition:] For each location $l_i$ that is monitored, the adversary adds a record in its knowledge base composed of the following fields:
 \begin{itemize}
  \item \emph{Location ID} (\emph{LocID} for brevity in the following): An identifier of the probed location $l_i\in\mathcal{L}$.
 \item \emph{Bytes}: The total size in bytes of the transmitted and received encrypted packets that belong to the same TLS/SSL session.
  \item \emph{Timestamp}: The timestamp of the first packet of the TLS/SSL session.
 \end{itemize}
\end{description}
\noindent
An example of knowledge base that the adversary would create is reported on the left of Table \ref{tab:featurematrix}.
\begin{table}[t]
 \centering
 \caption{Example of the adversary knowledge base, user dataset related to a user position during time, and possible guesses.}
 \begin{center}
  \begin{tabular}[t]{|r|l|l|}\hline
   \multicolumn{3}{|c|}{\textbf{Adversary Knowledge Base}}\\\hline
   \textbf{LocID} & \textbf{Bytes} & \textbf{Timestamp} \\\hline
    1 & 35780 & 1399743000 \\\hline
    2 & 30780 & 1399743000 \\\hline
    * & * & * \\\hline
    * & * & * \\\hline
    1 & 36780 & 1399743060 \\\hline
    2 & 30784 & 1399743060 \\\hline
  \end{tabular}
  \quad
  \begin{tabular}[t]{|r|l|l|}\hline
   \multicolumn{3}{|c|}{\textbf{User Dataset}}\\\hline
   \textbf{LocID} & \textbf{Bytes} & \textbf{Timestamp} \\\hline
    ? & 35780 & 1399743000 \\\hline
    ? & 35780 & 1399743020 \\\hline
    ? & 36780 & 1399743040 \\\hline
    ? & 36780 & 1399743060 \\\hline
    ? & 30784 & 1399743080 \\\hline
    ? & 30784 & 1399743100 \\\hline
  \end{tabular}
  \quad
 \end{center}
\label{tab:featurematrix}
\end{table} 

\subsection{Selection of the Candidate Locations}
\label{subsec:statistical}

To track a user's position, the adversary relies upon only two fields: the sum of the exchanged bytes of a TLS/SSL session and the timestamp. This information is derived from the header of the packets, which is not encrypted by the SSL protocol. 
To learn the position of a given user $U$ at time $t_0$, it is enough for the adversary to collect the communication traffic between the user's LBA and the LBS between time $t_0-t$ and $t_0$. For each TLS/SSL session, $\mathcal{A}$ calculates the fields described above and creates the user dataset. This is shown on the right of Table~\ref{tab:featurematrix}. \\
\indent
At any given moment, each location is potentially characterized by a fixed amount of bytes. As such, the adversary determines the candidate locations by analyzing those locations that have generated an amount of bytes similar to the entries of the user dataset. Namely, suppose that $\mathcal{A}$ wants to determine the position of the user $U$ at time $t_0$. $\mathcal{A}$ builds a filtered adversary knowledge base containing only the instances of the knowledge base such that their timestamps fall within the time frame $[t_0-t,t_0]$. The size of the time frame depends on the specific LBA and LBS taken into consideration, and on the typical behavior of the user. We assume that, within the targeted time frame, the user does not move from his location. The adversary restricts the number of possible locations studying the statistical distribution of the filtered adversary knowledge base and the user dataset. \\
\indent
In the experiments, we will consider also the case of time-misalignment between the filtered adversary knowledge base and user dataset. This case is useful when the adversary is not able to collect data during 
the time frame $[t_0-t,t_0]$. In such a case, the adversary may use a different time frame $[t_0-t-\delta,t_0-\delta]$, for a given $\delta>0$. \\
\indent
The adversary may use a statistical distance measure to quantify the distance between the samples of the user dataset, and the samples of the filtered adversary knowledge base, for each possible location. This will allow $\mathcal{A}$ to settle on a list of candidate locations where the user was at time $t_0$. \\
\indent
Several statistical distance measures can be used for this purpose. 
However, in the experimental section, we will show that our approach is accurate even when using a very simple distance metric. Let us indicate with $\bar{x}$ the user dataset, with $\bar{y}$ the filtered adversary knowledge base, and with $\mathcal{L}$ the set of all possible locations. Furthermore, with the notation $\bar{y}[l_i]$ we refer to the subset of the adversary knowledge base $\bar{y}$ related to an individual location $l_i \in \mathcal{L}$. In other words, $\bar{y}[l_i]$ contains all the instances of the filtered adversary knowledge base such that the field \emph{LocID} is equal to $l_i$. Then the adversary will select a candidate location set $\mathcal{S}$ of size $k$ in the following way:
\begin{equation}
\label{eq:cls}
\min_{\substack{\mathcal{S}\subseteq\mathcal{L}\\
                  \left\vert\mathcal{S}\right\vert=k}} \sum_{l_i\in \mathcal{S}} d(\bar{x},\bar{y}[l_i]) \;  ,
\end{equation}
\noindent
where $d(\bar{x},\bar{y}[l_i])$ indicates some statistical distance measure between $\bar{x}$ and $\bar{y}[l_i]$.
In the experiments, we used the following definition of distance: $d(\bar{x},\bar{y}[l_i])=\left\vert m(\bar{x})-m(\bar{y}[l_i])\right\vert$, where $m(\cdot)$ is the median function. Once the size $k$ is fixed,  Equation \ref{eq:cls} allows to select a candidate location set $\mathcal{S}$ of size $k$, composed of the locations $l_i$ that minimize the overall distance between $\bar{x}$ and $\bar{y}[l_i]$. 

\section{Experiments and Results}
\label{sec:exp-and-res}
To prove the feasibility and the accuracy of our approach, we performed a thorough analysis
of one of the most popular and advanced LBAs: GoogleNow. GoogleNow is an application provided by Google which comes preinstalled with the vast majority of Android devices~\cite{GoogleNow:Website}. It is also available on iOS, Google Glass and even on Android Wear devices. GoogleNow is not only a LBA, but it acts also as a personal assistant by providing personalized information to the user. User-based and location-based information are sent together within the same encrypted traffic. However, we will show that the use of encrypted communications does not hinder the process of identifying  user locations as long as some of the GoogleNow user's preferences are known a priori. \\
\indent
GoogleNow app operates regardless of the interactions with its users, and independently determines when information should be downloaded from the LBS server \cite{GoogleNow:AddRemoveCards} (unless a refresh is forced by the user). We can therefore speculate that GoogleNow app periodically sends the GPS location of the user to the LBS (Google servers). The LBS then replies with the information related to the GPS position sent by the GoogleNow app (e.g., nearby restaurants, bus stops, and/or images of the location). We confirmed this by installing the GoogleNow app on an Android X86 Virtual Machine (VM) and extracting the data exchanged between our VM and Google servers via a man-in-the-middle proxy. This experiment confirmed that the GoogleNow app sends the GPS coordinates of the user but also much more data, including the information currently displayed by the app on the smartphone. 

\subsection{Data Collection of GoogleNow Data} 
We performed a location spoofing attack to collect the encrypted network traffic exchanged by the GoogleNow app and the Google servers, and to build the adversary knowledge base as described in Sect.~\ref{subsec:datacollection}. Each spoofed GPS user location corresponds to a point the adversary is interested in monitoring. We used the mitm-proxy software for this task \cite{mitm} configured as a transparent proxy on a network of Androidx86 Virtual Machines run in a Virtualbox hypervisor on a Linux host. Each VM has been configured with a different user account. We configured different preferences for each account  to determine to what extent Google personalizes the data sent to each user. Because of space restrictions, we report on the results achieved for a single configuration of the user preferences, that is the one where the user settled its home and work locations only. Similar results were achieved for the other accounts that we analyzed and will be reported in the full version of this paper. \\
\indent
GPS locations exchanged between the app and the server are encrypted within SSL, thus we performed a man-in-the-middle attack. To this aim, a self-signed certificate was copied into each VM and added to the Android's root certificates. In this way, the SSL traffic was decrypted and we found out that GoogleNow uses a protocol called Protobuf as data interchange format ~\cite{protobuf}. This protocol was designed by Google to be smaller and faster than XML. By analyzing the data structure of the protobuf messages that were intercepted by the mitm-proxy, we were able to identify the fields that contain the latitude and longitude of the user. These fields are sent to the Google servers in all the HTTPS requests that contains the following string within the URL: ``tg/fe/request". \\
\indent
We collected data about an area of two square kilometers positioned in the center of a large European city. The adversary simulated the presence of its dummy users in the area by moving them on a grid of $5\times10$ points ($50$ different \emph{LocID} in total). The location (i,j) of the grid is identified by the label \verb i_j . Sect.~\ref{sec:areasize} provides  the motivation behind the selection of these parameters. In order to collect a large amount of data in a short time range, rather than changing the user position of the VM through a MockLocationProvider and then wait for a GoogleNow genuine request, the adversary replayed a request containing the ``tg/fe/request'' string in the URL several times, modifying the actual position with every point of the grid. The network traffic intercepted by the proxy has been used to feed the knowledge base of the adversary. \\
\indent
For the experiments we had to simulate the target users as well. Therefore, during the same period of three weeks, different VMs have been configured with distinct user accounts. Users moved into the monitored area, and their respective traffic was collected and stored separately. Clearly, this last step will not be performed when tracing real users.

\subsection{Exploratory Data Analysis}
\label{subsec:exploratory}

\begin{figure}[t]
\centering
 \epsfig{file=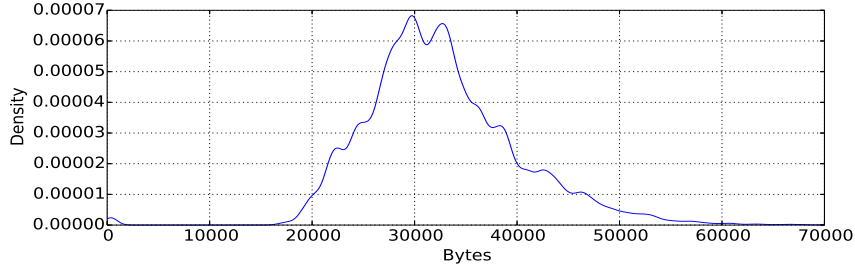,height=1.5in,width=5in}
\caption{Probability Distribution of the Bytes exchanged between the GoogleNow app and the Google servers.}
\label{fig:totalDensity}
\end{figure}

In this section, we analyze the collected data to improve our understanding of the GoogleNow traffic dataset. 
This exploratory data analysis is performed over the data collected for an account that has been configured by only specifying home and work locations. All the remaining user profiles that we analyzed show a very similar behavior. Figure~\ref{fig:totalDensity} reports on the probability distribution of the bytes per SSL session exchanged between the GoogleNow app and the Google servers during the entire monitored period. 
On average, $32{,}604$ bytes were exchanged per session, with a standard deviation of $7{,}518$. The minimum recorded value is equal to $80$, while the maximum is equal to $83{,}831$ bytes. The median is $31{,}804$ bytes, while lower and upper quartiles are equal to $27{,}791$ and $36{,}520$, respectively.

\begin{figure}[t]
\centering
\begin{subfigure}{\linewidth}
 \epsfig{file=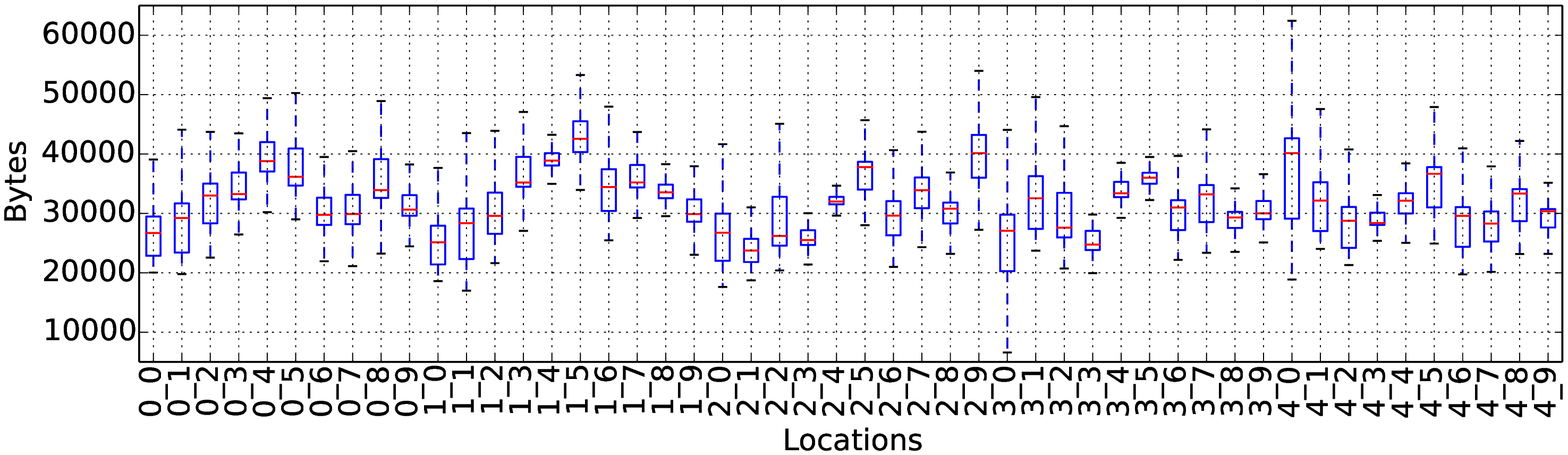,height=1.5in,width=5in}
\caption{Analysis of the entire adversary knowledge base (3-weeks of traffic).}
\label{fig:boxplotPerClass}
\end{subfigure}
\begin{subfigure}{\linewidth}
 \epsfig{file=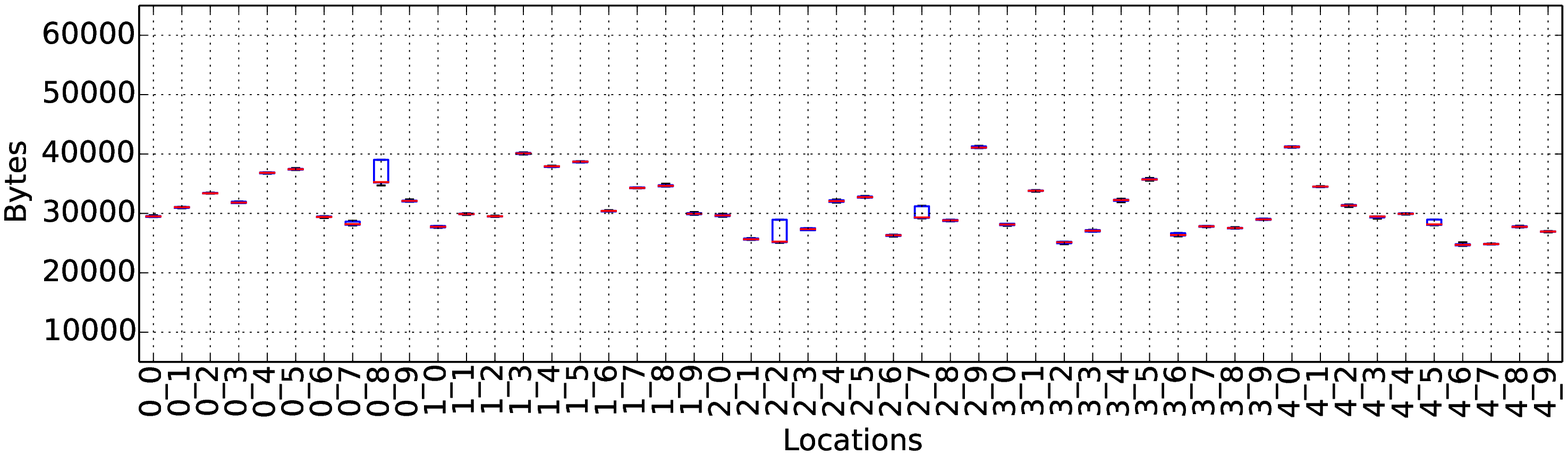,height=1.5in,width=5in}
\caption{Analysis of only a random hour of the adversary knowledge base.}
\label{fig:boxplotPerClass1hour}
\end{subfigure}
\caption{Statistical distribution of the bytes exchanged between the GoogleNow app and the Google servers per monitored location. }
\label{fig:boxplot}
\end{figure}

\indent
To determine whether the bytes exchanged between the GoogleNow app and the Google servers might be useful to identify the actual location of the user, we analyzed the statistical distribution of the bytes exchanged in each monitored location. Figure~\ref{fig:boxplot} shows the boxplots diagram for all the $50$ locations of the adversary knowledge base. The boxes extend from the lower to the upper quartile values of the data, with a line at the median. The whiskers extend from the boxes indicating variability outside the lower and upper quartiles. Figure~\ref{fig:boxplotPerClass} shows the statistical distribution of the bytes received in all the locations during the entire period of collection (3 weeks). Figure~\ref{fig:boxplotPerClass1hour} was obtained while analyzing only one hour of traffic randomly selected among the three weeks. Note that in Fig.~\ref{fig:boxplotPerClass}, all the locations show a similar behavior. They have a mean value that is rarely  greater than $40{,}000$ and lesser than $20{,}000$. The variance is very high, and lower and upper quartiles are quite far from the median value. With statistical distributions so similar to each other, it might be difficult to infer the actual location of a user. We quickly realized that the time of the day has a great influence on the information provided by GoogleNow, thus we limited the analysis to a period of time one hour long, randomly selected among the three weeks period. Figure~\ref{fig:boxplotPerClass1hour} shows the boxplot diagram related to this subset of data. It can be observed that almost all the locations have a very tight variance. First and third quartile are very close to each other. Furthermore, once a particular size is selected, only a few locations may have produced it. This is the main reason that led us to consider the time as an important parameter in our analysis.

\paragraph{Daily Pattern.}
During the exploratory data analysis, we realized that the amount of bytes exchanged between GoogleNow and the Google servers follows a daily pattern distribution. In particular, during daily hours it ranges from $26{,}000$ to $32{,}000$ bytes, whereas during the night it falls down in the interval between $22{,}000$ and $24{,}000$ bytes. In Sect.~\ref{subsec:accuracresult}, we will show how this daily pattern influences the accuracy results.

\subsection{Accuracy Results}
\label{subsec:accuracresult}
\begin{figure}[t]
\centering
\epsfig{file=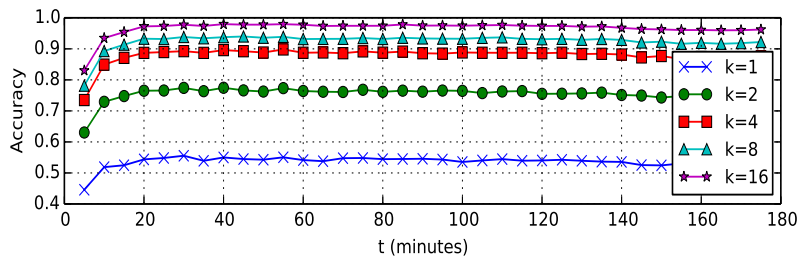,height=1.5in,width=5in}
\caption{Accuracy of Locations Sets: effect upon accuracy of varying $k$ (size of the candidate locations set) and $t$ (size of the time frame). }
\label{fig:accuracyCandidateLocations}
\end{figure}

\begin{figure}[t]
\centering
\includegraphics[scale=0.6]{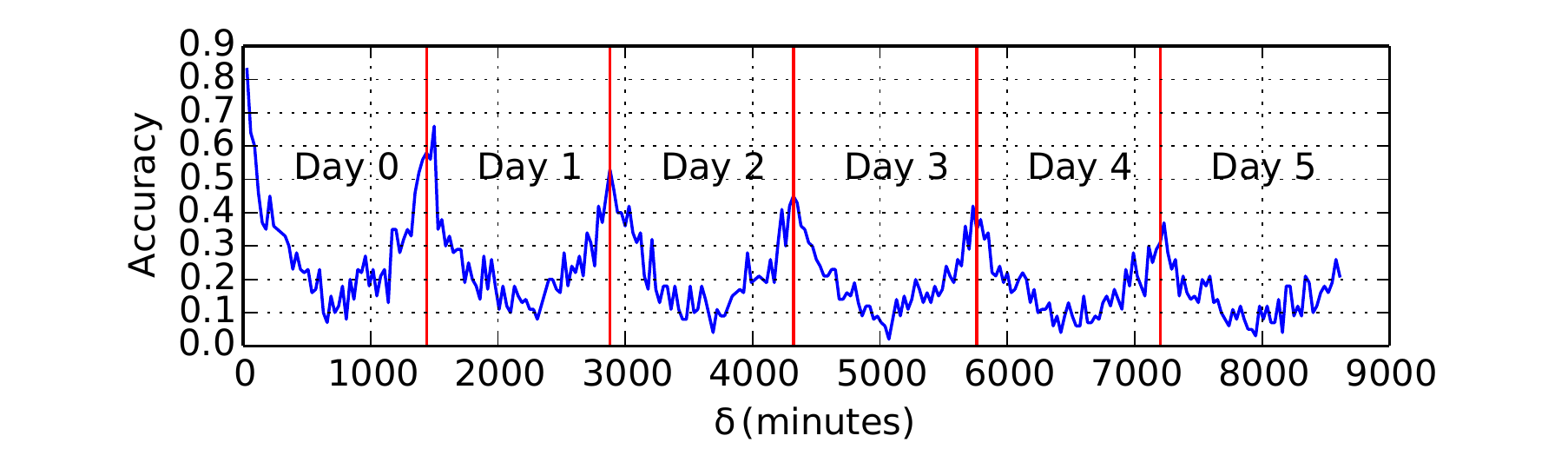}
\caption{Accuracy of Locations Sets: Effect upon accuracy of varying $\delta$ (time delay between the filtered adversary knowledge base and the user dataset). ($k=4$, $t=60$) }
\label{fig:accuracyCandidateLocationsDelta}
\end{figure}

In the following, we will report on the results of the tests that we performed on the collected dataset. All the experiments presented next represent the average of $10{,}000$ tests. In each test, the adversary tries to infer the position of a user. The $k$-identifiability of a tested user position is defined as 1 if the actual position of the user is within the set $\mathcal{S}$ of $k$ candidate locations selected with the approach described in Sect.~\ref{subsec:statistical}. Otherwise $k$-identifiability is defined as 0. Thus, the $k$-accuracy is the average of the $k$-identifiability values of each tested instance \cite{Liberatore:2006:ISE:1180405.1180437}. \\
\indent
In Fig.~\ref{fig:accuracyCandidateLocations}, we show the $k$-accuracy of a user location when varying $k$ and $t$, where $k$ is the size of the candidate locations set and $t$ is the size of time frame used to filter the adversary knowledge base. Observe that for $k=8$, we reach a value close to $95\%$ with a time frame of only $20$ minutes. The behavior of the $k$-accuracy is asymptotic and it reaches a value close to the maximum already at around $t=20$. In other words, the adversary has to analyze only $20$ minutes of traffic to reach the best accuracy performance, independently of the value $k$ selected. Other combinations of $k$ and $t$ also provide reasonable accuracy performance. For instance, for  $k=8$, $5$ minutes of traffic are enough to reach an accuracy of $79\%$, which is quite remarkable. \\
\indent
In Fig.~\ref{fig:accuracyCandidateLocationsDelta}, we show the effect of varying $\delta$, that is the difference of time between the filtered adversary knowledge base and the user dataset. In general, larger delays result in lower accuracy. However, the figure shows a cyclic behavior that reflects the daily activity highlighted in Sect.~\ref{subsec:exploratory}. 
Thus, if the adversary does not have in its knowledge base instances that fall in the same time frame of the user dataset, then it is better to use a filtered knowledge base that is one day older (1440 minutes) than one that is only 12 hours older (720 minutes). Indeed, in the former case the accuracy is slightly below $60\%$, while in the latter case it is around $12\%$ only. The figure also shows a decrease of the peaks that are in correspondence of every 24 hours. This is mainly due to the fact that the information provided by the app is being constantly updated, and it becomes obsolete after a few days.

\begin{figure}[htbp]
\centering
  \begin{subfigure}[b]{0.49\textwidth}
    \includegraphics[width=5.7cm,height=3.25cm]{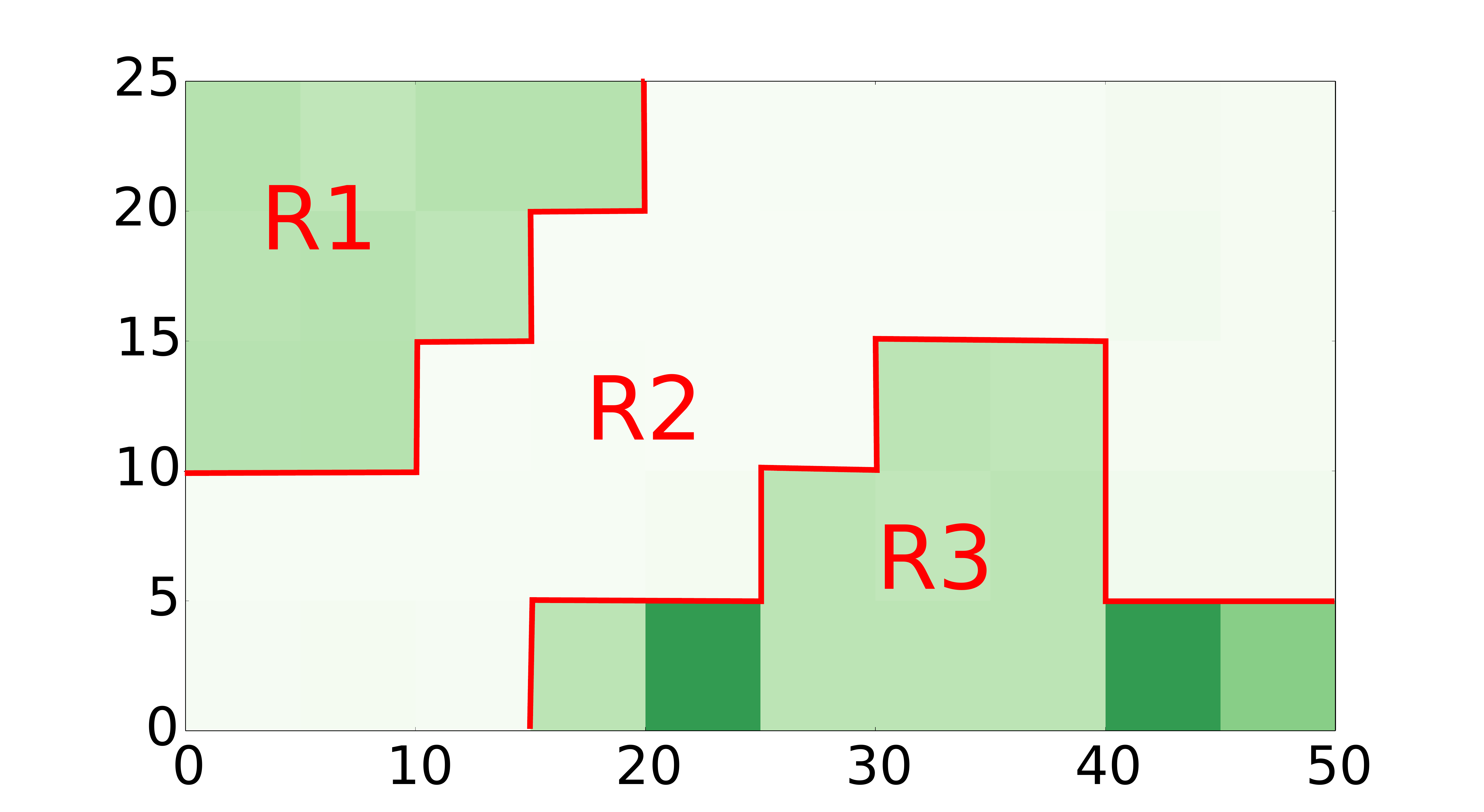}
    \caption{5 meters cells}
    \label{fig:5mgrid}
  \end{subfigure}
  \begin{subfigure}[b]{0.49\textwidth}
    \includegraphics[width=\textwidth]{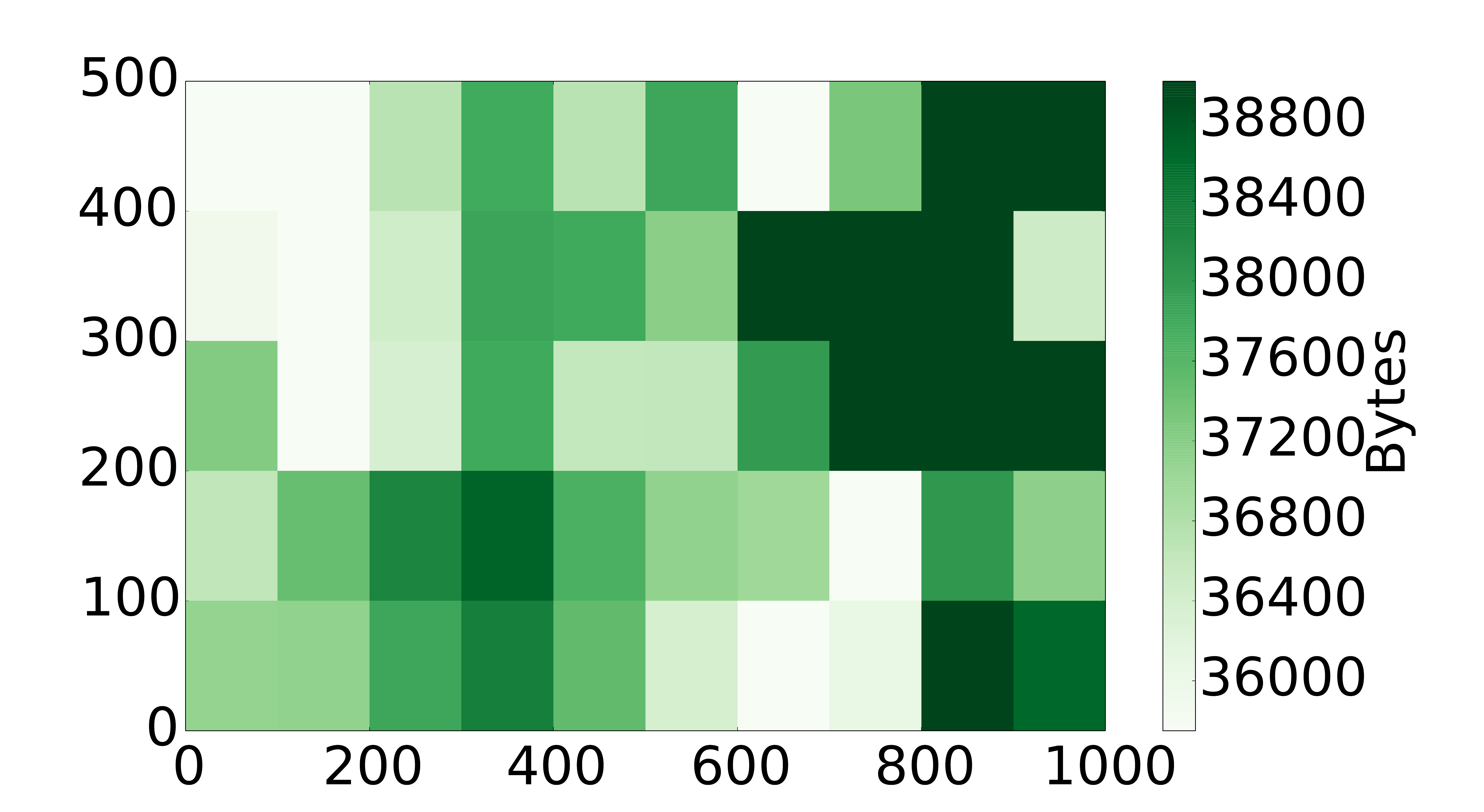}
    \caption{100 meters cells}
    \label{fig:100mgrid}
  \end{subfigure}  
  \caption{ Heat-matrices for different sizes of the cells}
  \label{fig:heatmatrices}
\end{figure}

\subsection{Granularity of the Monitored Area}
\label{sec:areasize}
In the experiments that we reported in Sect.~\ref{sec:exp-and-res}, the adversary monitored $50$ points that were distributed within an area of two square kilometers. Each point represented therefore a square area of 200 $m^2$. The adversary arbitrarily chooses one point within this square as a representative. However, the granularity of 200 $m^2$ does not necessarily match the granularity adopted by GoogleNow: if the granularity of GoogleNow is finer, then there are points within the area that have not been considered during the collection phase; on the other hand, if the granularity of GoogleNow is coarser, then several different points may become indistinguishable, thus impeding detection. \\
\indent
We performed an additional experiment to learn the granularity used by Google services in a given area. To this aim, we run the data collection phase using different granularities: 5, 10, 25 and 100 $m^2$. The number of points is fixed to 50 in all cases considered. Each square was probed once every 5 minutes and we calculated the median of the exchanged bytes. Figure~\ref{fig:heatmatrices} shows the results achieved for the two extreme values that we used: 5 and 100 $m^2$. In Fig.~\ref{fig:5mgrid}, it is possible to easily identify three different regions with a similar amount of bytes exchanged (they are indicated with $R1$, $R2$, $R3$). In Fig.~\ref{fig:100mgrid}, a cell represents an area of 100 $m^2$. Even in this case there exist regions containing more than one indistinguishable square. The shapes of these regions are irregular since they depend on the location, on the importance, and on the number of the points of interests that fall in those areas. \\
\indent
The result of this analysis is that the granularity of GoogleNow in the particular region we analyzed is dynamic and ranges from  5 $m^2$ to 100 $m^2$.  Thus the most appropriate granularity should be selected to find the right balance between performance (data collection) and detection accuracy. Our experiments were run with 200 $m^2$ in our target region because the detection accuracy was very high regardless and  this allowed us to speed up data collection (the entire area could be monitored every 5 minutes) and avoid overloading GoogleNow servers with our requests.  

\section{Related Work}
\label{sec:related}
The closest research areas to this work, are traffic analysis and location obfuscation. In the following, both of them will be briefly described. \\
\indent
\textit{Traffic Analysis} is devoted to exploiting observable features in an encrypted traffic to infer information about the content of the communication. For instance,~\cite{Berthold:2000:PLU:332186.332211,Raymond:2001:TAP:371931.371972} leverage observables such as the timing and the exchanged bytes to discover communication patterns that can be used to break the anonymity or the confidentiality of the communication. The majority of the work in this area has been conducted over HTTPS protocol~\cite{Liberatore:2006:ISE:1180405.1180437,Herrmann:2009:WFA:1655008.1655013,Panchenko:2011:WFO:2046556.2046570,Luo11httpos:sealing}, even though other protocols such as VoIP \cite{Wright:2008:SMY:1397759.1398055} have been analyzed as well. A variety of techniques such as Naive Bayes classifiers, Jaccard's coefficient \cite{Liberatore:2006:ISE:1180405.1180437}, common text mining techniques applied to the normalized frequency distribution of observable IP packets \cite{Herrmann:2009:WFA:1655008.1655013} or vector machine classifier \cite{Panchenko:2011:WFO:2046556.2046570} have been adopted and applied to traffic analysis in order to identify which website the target user had accessed. These techniques work even if the communication is encrypted or anonymized through networks such as Tor. Indeed, \cite{Panchenko:2011:WFO:2046556.2046570} achieved an astonishing accuracy of 97\% in these cases. Several countermeasures have also been devised~\cite{Luo11httpos:sealing,Liberatore:2006:ISE:1180405.1180437,Wright09trafficmorphing}. They work by manipulating packet size, web object size, flow size, and the timing of the packets to hinder  traffic analysis. Unfortunately, these countermeasures have significant performance drawbacks \cite{Liberatore:2006:ISE:1180405.1180437} and some of them are particularly vulnerable to simple attacks \cite{Dyer:2012:PIS:2310656.2310689} that exploit the coarse features of traffic (e.g., total time and bandwidth). \\
\indent
\textit{Location Obfuscation} aims at hiding the exact geographical location of the user from the LBSs. Considering both location information accuracy and privacy,~\cite{Samarati} introduces the concept of relevance which is used to protect the location information of users together with other obfuscation operators. In~\cite{Gruteser:2003:AUL:1066116.1189037}, the authors propose to adjust the resolution of location information along spatial or temporal dimensions to meet specific anonymity constraints. In~\cite{Chow:2006:PSC:1183471.1183500}, a peer-to-peer method is proposed where users cooperate to hide their real location from the LBS. Another approach proposed in~\cite{Andres:2013:GDP:2541806.2516735} consists of adding controlled noise to the user's location to obtain an approximate version of it which is then sent to the LBS. An obfuscation technique is proposed in \cite{DBLP:journals/ton/RiboniVVBM15} for a number of sensitive data, including IP addresses of users, this tecnique provides formal confidentiality guarantees under realistic assumptions about the adversary's knowledge.

\section{Conclusions}
\label{sec:conclusions}
In this paper, we introduced a new adversary model for Location Based Services. The model takes into account 
an unauthorized third party, different from the LBS provider itself, that wants to infer the position of a
LBS user. We analyzed one of the most popular location based apps available on Android, that is GoogleNow, and we shown that our adversary can infer the position of a user with an accuracy of almost $90\%$ through a statistical analysis of the user's encrypted network traffic. 

\section*{Acknowledgements}
\label{sec:ack}
This work has been partially supported by the European Commission H2020 under the SUNFISH project, N.644666, and by the European Commission Directorate General Home Affairs, under the GAINS project, HOME/2013/CIPS/AG/4000005057. \\
\\
The final publication is available at \url{link.springer.com}.

\bibliographystyle{plain}
\bibliography{gnow-main}

\begin{thebibliography}{10}

\bibitem{mitm}
Man in the middle proxy.
\newblock \url{https://mitmproxy.org/}.

\bibitem{protobuf}
Protocol buffers - google's data interchange format.
\newblock \url{https://github.com/google/protobuf}, 2008.

\bibitem{ArsTechnica:MachinesStealPhoneData}
Meet the machines that steal your phone's data | ars technica.
\newblock \url{http://tinyurl.com/o9vd4u9}, 2013.

\bibitem{NSA:TorAttack}
Schneier on security: How the nsa attacks tor/firefox users with quantum and
  foxacid.
\newblock \url{http://tinyurl.com/n84axpz}, 2013.

\bibitem{WashingtonPost:Systems}
For sale: Systems that can secretly track where cellphone users go around the
  globe - the washington post.
\newblock \url{http://tinyurl.com/kuazdjs}, 2014.

\bibitem{CMU:Raffalink}
Your location has been shared 5398 times.
\newblock \url{http://tinyurl.com/nuh6w4e}, 2015.

\bibitem{Andres:2013:GDP:2541806.2516735}
M.~E. Andr{\'e}s, N.~E. Bordenabe, K.~Chatzikokolakis, and C.~Palamidessi.
\newblock Geo-indistinguishability: differential privacy for location-based
  systems.
\newblock In {\em Proc. of the 2013 ACM SIGSAC conference on Computer and
  communications security}, CCS '13, pages 901--914, New York, NY, USA, 2013.
  ACM.

\bibitem{Samarati}
C.A. Ardagna, M.~Cremonini, S.~De~Capitani~di Vimercati, and P.~Samarati.
\newblock An obfuscation-based approach for protecting location privacy.
\newblock {\em IEEE Transactions on Dependable and Secure Computing},
  8(1):13--27, Jan 2011.

\bibitem{Berthold:2000:PLU:332186.332211}
O.~Berthold, H.~Federrath, and M.~K\"{o}hntopp.
\newblock Project anonymity and unobservability in the internet.
\newblock In {\em Proc. of the Tenth Conference on Computers, Freedom and
  Privacy: Challenging the Assumptions}, CFP '00, pages 57--65, New York, NY,
  USA, 2000. ACM.

\bibitem{Chow:2006:PSC:1183471.1183500}
C.-Y. Chow, M.~F. Mokbel, and X.~Liu.
\newblock A peer-to-peer spatial cloaking algorithm for anonymous
  location-based service.
\newblock In {\em Proc. of the 14th Annual ACM International Symposium on
  Advances in Geographic Information Systems}, GIS '06, pages 171--178, New
  York, NY, USA, 2006. ACM.

\bibitem{Conti:2015:CYH:2699026.2699119}
M.~Conti, L.~V. Mancini, R.~Spolaor, and N.~V. Verde.
\newblock Can't you hear me knocking: Identification of user actions on android
  apps via traffic analysis.
\newblock In {\em Proceedings of the 5th ACM Conference on Data and Application
  Security and Privacy}, CODASPY '15, pages 297--304, New York, NY, USA, 2015.
  ACM.

\bibitem{Dyer:2012:PIS:2310656.2310689}
K.~P. Dyer, S.~E. Coull, T.~Ristenpart, and T.~Shrimpton.
\newblock Peek-a-boo, i still see you: Why efficient traffic analysis
  countermeasures fail.
\newblock In {\em Proc. of the 2012 IEEE Symposium on Security and Privacy}, SP
  '12, pages 332--346, Washington, DC, USA, 2012. IEEE Computer Society.

\bibitem{GoogleNow:AddRemoveCards}
Google.com.
\newblock Add or remove now cards.
\newblock \url{http://tinyurl.com/ppy4svc}, 2015.

\bibitem{GoogleNow:Website}
Google.com.
\newblock Google now.
\newblock \url{https://www.google.com/landing/now/}, 2015.

\bibitem{Gruteser:2003:AUL:1066116.1189037}
M.~Gruteser and D.~Grunwald.
\newblock Anonymous usage of location-based services through spatial and
  temporal cloaking.
\newblock In {\em Proc. of the 1st International Conference on Mobile Systems,
  Applications and Services}, MobiSys '03, pages 31--42, New York, NY, USA,
  2003. ACM.

\bibitem{Herrmann:2009:WFA:1655008.1655013}
D.~Herrmann, R.~Wendolsky, and H.~Federrath.
\newblock Website fingerprinting: Attacking popular privacy enhancing
  technologies with the multinomial naive-bayes classifier.
\newblock In {\em Proc. of the 2009 ACM Workshop on Cloud Computing Security},
  CCSW '09, pages 31--42, New York, NY, USA, 2009. ACM.

\bibitem{Liberatore:2006:ISE:1180405.1180437}
M.~Liberatore and B.~N. Levine.
\newblock Inferring the source of encrypted http connections.
\newblock In {\em Proc. of the 13th ACM Conference on Computer and
  Communications Security}, New York, NY, USA, 2006. ACM.

\bibitem{Luo11httpos:sealing}
X.~Luo, P.~Zhou, E.~W.~W. Chan, W.~Lee, R.~K.~C. Chang, and R.~Perdisci.
\newblock Httpos: Sealing information leaks with browser-side obfuscation of
  encrypted flows.
\newblock In {\em In Proc. Network and Distributed Systems Symposium (NDSS).
  The Internet Society}, 2011.

\bibitem{Panchenko:2011:WFO:2046556.2046570}
A.~Panchenko, L.~Niessen, A.~Zinnen, and T.~Engel.
\newblock Website fingerprinting in onion routing based anonymization networks.
\newblock In {\em Proc. of the 10th Annual ACM Workshop on Privacy in the
  Electronic Society}, WPES '11, pages 103--114, New York, NY, USA, 2011. ACM.

\bibitem{Raymond:2001:TAP:371931.371972}
J.~Raymond.
\newblock Traffic analysis: Protocols, attacks, design issues, and open
  problems.
\newblock In {\em International Workshop on Designing Privacy Enhancing
  Technologies: Design Issues in Anonymity and Unobservability}, pages 10--29,
  New York, NY, USA, 2001. Springer-Verlag New York, Inc.

\bibitem{DBLP:journals/ton/RiboniVVBM15}
D.~Riboni, A.~Villani, D.~Vitali, C.~Bettini, and L.~V. Mancini.
\newblock Obfuscation of sensitive data for incremental release of network
  flows.
\newblock {\em {IEEE/ACM} Transactions on Networking}, 23(2):672--686, 2015.

\bibitem{Stober:2013:YSY:2462096.2462099}
T.~St\"{o}ber, M.~Frank, J.~Schmitt, and I.~Martinovic.
\newblock Who do you sync you are?: Smartphone fingerprinting via application
  behaviour.
\newblock In {\em Proc. of ACM WiSec}, 2013.

\bibitem{NinoVerdeNATleftBehind}
N.~V. Verde, G.~Ateniese, E.~Gabrielli, L.~V. Mancini, and A.~Spognardi.
\newblock No nat'd user left behind: Fingerprinting users behind nat from
  netflow records alone.
\newblock In {\em Proc. of the 2014 IEEE 34th International Conference on
  Distributed Computing Systems}, ICDCS '14, pages 218--227, Madrid, Spain,
  2014. IEEE Computer Society.

\bibitem{Wright:2008:SMY:1397759.1398055}
C.~V. Wright, L.~Ballard, S.~E. Coull, F.~Monrose, and G.~M. Masson.
\newblock Spot me if you can: Uncovering spoken phrases in encrypted voip
  conversations.
\newblock In {\em Proc. of the 2008 IEEE Symposium on Security and Privacy}, SP
  '08, pages 35--49, Washington, DC, USA, 2008. IEEE Computer Society.

\bibitem{Wright09trafficmorphing}
C.~V. Wright, S.~E. Coull, and F.~Monrose.
\newblock Traffic morphing: An efficient defense against statistical traffic
  analysis.
\newblock In {\em In Proc. of the 16th Network and Distributed Security
  Symposium}, pages 237--250. IEEE, 2009.

\end{thebibliography}

\end{document}